\documentclass{article}

\PassOptionsToPackage{numbers, sort&compress}{natbib}

\usepackage[preprint]{neurips_2026}

\usepackage[utf8]{inputenc} 
\usepackage[T1]{fontenc}    
\usepackage{hyperref}       
\usepackage{url}            
\usepackage{booktabs}       
\usepackage{amsmath}        
\usepackage{amsfonts}       
\usepackage{microtype}      
\usepackage{xcolor}         
\usepackage{tikz}           
\usepackage{graphicx}       
\usepackage{subcaption}     
\usepackage{placeins}       
\usetikzlibrary{arrows.meta, positioning, calc}
\title{Knowing When Not to Answer: Pseudo-Ensembles\\ for Abstention in Music Audio-Language Models}

\author{%
  Aanya Maheshwari\thanks{Correspondence to: \texttt{aanya.m1\_jcd@gemsed.com}} \\
  Jumeirah College, Dubai \\
  \And
  Vatsal Raina \\
  Apta AI, Spark AI Research \\
}

\begin{document}

\maketitle

\begin{abstract}
Music audio-language models are evaluated almost entirely by accuracy on multiple-choice questions.
This protocol forces the model to commit to an option, so a lucky guess looks the same as real
musical understanding. What is missing is a way to tell when the model does not know the answer, so
that it can abstain instead of guessing. The usual solution, an ensemble of independently trained
models, is far too expensive here, which leaves the entropy of a single predictive distribution as
the only available confidence signal. We instead build \emph{pseudo-ensembles} from one pretrained
model by perturbing its input in ways that cannot change the correct answer, then averaging the
resulting distributions over the options. Our main construction simply shuffles the order in which
the candidate answers are presented; we also study ensembles built from corrupted audio and from
swapped option labels. A pseudo-ensemble gives several predictive distributions per question, so it
supports the full family of ensemble-based uncertainty measures (entropy of the expected
distribution, expected entropy, and their difference, the mutual information) rather than entropy
alone. Evaluating TinyMU on MuChoMusic, we find that averaging over four option orderings raises
accuracy from $55.7\%$ to $59.2\%$, and that the resulting uncertainty measures rank the
model's errors better than the single-pass entropy baseline, reducing the area under the
error retention curve from $0.293$ to $0.261$. All of this costs a few extra forward
passes and no retraining, which makes abstention practical for compact music
audio-language models.
\end{abstract}

\section{Introduction}

Large language models have been extended with audio encoders that can listen to a recording and
answer questions about it in natural
language~\citep{gong2024ltu,tang2024salmonn,deshmukh2023pengi,chu2024qwen2audio,kong2024audioflamingo},
and a growing number of these systems target music
specifically~\citep{liu2024mullama,gardner2024llark,deng2024musilingo,ma2024foundation}. Compact
models such as TinyMU ($229$M parameters) now reach much of the accuracy of far larger
systems~\citep{li2026tinymu}. Free-form musical descriptions are hard to score, so evaluation has
settled on multiple-choice question answering (MCQA), with benchmarks such as MuChoMusic asking a
model to pick one of four answers about a short excerpt~\citep{weck2024muchomusic}. The protocol
forces a commitment, so a lucky guess looks the same as real musical understanding. Published
evaluations report accuracy and little else: we know how often these models are right, not whether
they know when they are wrong.

That gap matters. A music assistant that can abstain or defer when unsure would make the answers it
does give more trustworthy, and would expose the difference between a model that guesses well and
a model that understands music. This is selective
prediction~\citep{elyaniv2010foundations,geifman2017selective}: the system needs a scalar
uncertainty score by which to rank predictions. Softmax confidence and entropy are the usual
single-model scores~\citep{hendrycks2017baseline}, but they are poorly
calibrated~\citep{guo2017calibration}. Ensembles of independently trained models give a richer
signal~\citep{lakshminarayanan2017simple,gal2016dropout}, decomposing predictive uncertainty into a
\emph{total} term (entropy of the average), a \emph{data} term (average member entropy) and their
difference, the mutual
information~\citep{depeweg2018decomposition,malinin2018predictive,malinin2021uncertainty}. Training
such an ensemble of music audio-language models is far too expensive. Work on text-only LLMs has
turned to self-reported confidence~\citep{kadavath2022know,lin2022teaching,xiong2024can}, semantic
consistency~\citep{kuhn2023semantic,farquhar2024detecting} and cross-model
agreement~\citep{feng2024abstain}, but these target open-ended generation, and none have been
studied on audio or music. Meanwhile MCQA itself is fragile: language models show systematic
position
biases~\citep{robinson2023mcp,zheng2024selectors,pezeshkpour2024sensitivity,alzahrani2024benchmarks}.
That fragility, usually treated as a nuisance, is exactly what makes cheap uncertainty estimation
possible.

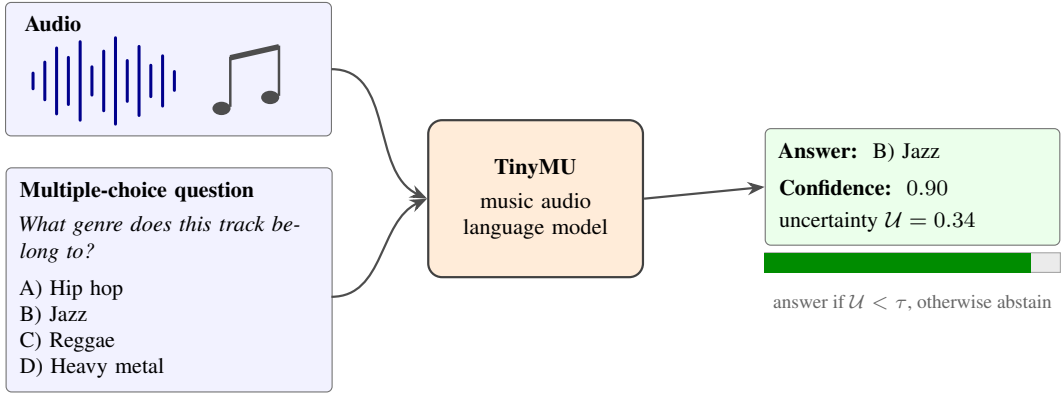
\begin{figure}[t]
  \centering
  \resizebox{\linewidth}{!}{\begin{tikzpicture}[
      font=\footnotesize,
      >={Stealth[length=2.2mm, width=1.7mm]},
      inbox/.style  = {draw=black!55, rounded corners=2pt, fill=blue!5},
      modelbox/.style = {draw=black!75, thick, rounded corners=5pt, fill=orange!15},
      outbox/.style = {draw=black!55, rounded corners=2pt, fill=green!8},
    ]

    \draw[inbox] (0,1.85) rectangle (4.15,3.55);
    \node[anchor=north west, font=\footnotesize\bfseries] at (0.12,3.50) {Audio};
    \foreach \i/\h in {0/0.10, 1/0.24, 2/0.42, 3/0.30, 4/0.50, 5/0.18,
                       6/0.38, 7/0.55, 8/0.26, 9/0.44, 10/0.20, 11/0.32, 12/0.12}
      \draw[line width=1.1pt, blue!55!black, line cap=round]
        ({0.35+\i*0.15}, {2.55-\h}) -- ({0.35+\i*0.15}, {2.55+\h});
    \begin{scope}[shift={(2.75,2.20)}, black!70]
      \fill (0,0)    ellipse (0.115 and 0.085);
      \fill (0.62,0.14) ellipse (0.115 and 0.085);
      \draw[line width=0.9pt] (0.112,0.02)  -- (0.112,0.66);
      \draw[line width=0.9pt] (0.732,0.16) -- (0.732,0.80);
      \draw[line width=2.0pt] (0.112,0.645) -- (0.732,0.785);
    \end{scope}

    \node[inbox, anchor=north west, text width=3.8cm, align=left, inner sep=5pt]
      at (0,1.45) {\textbf{Multiple-choice question}\\[3pt]
        \emph{What genre does this track belong to?}\\[4pt]
        A) Hip hop\\ B) Jazz\\ C) Reggae\\ D) Heavy metal};

    \node[modelbox, text width=2.5cm, align=center, minimum height=2.0cm]
      (model) at (6.75,1.05) {\textbf{TinyMU}\\[3pt] music audio\\ language model};

    \node[outbox, text width=3.4cm, align=left, inner sep=5pt]
      (out) at (11.55,1.20) {\textbf{Answer:}~~B) Jazz\\[4pt]
        \textbf{Confidence:}~~$0.90$\\[2pt]
        uncertainty $\mathcal{U}=0.34$};

    \draw[fill=black!8, draw=black!35]
      ($(out.south west)+(0,-0.34)$) rectangle ($(out.south east)+(0,-0.09)$);
    \draw[fill=green!55!black, draw=none]
      ($(out.south west)+(0,-0.34)$) rectangle
      ($(out.south west)!0.9!(out.south east)+(0,-0.09)$);

    \draw[->, thick, black!70] (4.15,2.70) to[out=0, in=155] (model.west);
    \draw[->, thick, black!70] (4.15,-0.20) to[out=0, in=205] (model.west);
    \draw[->, thick, black!70] (model.east) -- (out.west);

    \node[anchor=north, font=\scriptsize, text=black!60] at ($(out.south)+(0,-0.50)$)
      {answer if $\mathcal{U}<\tau$, otherwise abstain};

  \end{tikzpicture}}
  \caption{Overview of the task. The model returns both an answer and an uncertainty score, and
  abstains when the uncertainty exceeds a threshold $\tau$.}
  \label{fig:overview}
\end{figure}

This work studies abstention for music audio-language models on multiple-choice musical question
answering (Figure~\ref{fig:overview}). We treat MCQA as classification over the option symbols,
reading each option's probability off the first-token distribution, and then build a
\emph{pseudo-ensemble} from a single pretrained model by perturbing its input in ways that cannot
change the correct answer. The construction we focus on shuffles the order of the candidate
answers; we also study corrupted audio and swapped option labels. Averaging the resulting
distributions makes the standard ensemble measures available, entropy of the expected distribution,
expected entropy and mutual information, alongside negative confidence, for a few forward passes
rather than a training run. We apply this to TinyMU~\citep{li2026tinymu} on
MuChoMusic~\citep{weck2024muchomusic} and assess each measure with error retention curves, which
track how fast the error rate falls as the most uncertain questions are handed over.

Our contributions are as follows.
\begin{itemize}\itemsep 1pt
  \item We frame music audio-language evaluation as selective prediction, and present the first
        study of uncertainty quantification and abstention for a music audio-language model.
  \item We introduce \emph{pseudo-ensembles}: ensembles from a single pretrained model via
        answer-preserving input perturbations, unlocking ensemble uncertainty measures at the cost
        of a few extra forward passes and no retraining.
  \item We show that shuffling the candidate answers both raises accuracy and ranks the model's
        errors better than the entropy of a single pass.
\end{itemize}

\section{Uncertainty quantification for classification}
\label{sec:uq}

Answering a multiple-choice question is a classification problem: the model assigns a probability
to each of the $K$ candidate answers and returns the most probable one. Two predictions with the
same $\arg\max$ can still differ sharply in how concentrated their belief is. Uncertainty
quantification supplies a scalar that is large when belief is spread and small when it is
concentrated. This section sets out the measures we use; how we obtain the distributions is the
subject of Sections~\ref{sec:mcqa} and~\ref{sec:abstention}.

Write $x$ for a complete input, meaning the audio excerpt, the question and the $K$ candidate
answers, and $y \in \{1,\dots,K\}$ for the index of the chosen option, with $K=4$ throughout. A
model with parameters $\theta$ defines a categorical distribution $p$ over the options, with
$p_k = P(y = k \mid x, \theta)$, and predicts $\hat{y} = \arg\max_k p_k$.

\paragraph{Measures from a single distribution.}
The simplest score is the probability of the chosen option, $\max_k p_k$. To keep the convention
that larger always means less reliable we use \emph{negative confidence}, $-\max_k p_k$. It ignores
the rest of the distribution, so it cannot tell a model torn between two options from one spreading
doubt over three. \emph{Entropy} uses the whole distribution,
\begin{equation}
  \mathcal{H}[p] \;=\; -\sum_{k=1}^{K} p_k \log p_k \;\in\; \left[0,\, \log K\right] ,
  \label{eq:entropy}
\end{equation}
Entropy is zero when one option takes all the mass and reaches $\log K \approx 1.39$ nats for
$K=4$ when the model guesses uniformly. Both scores collapse two situations that call for
different responses: a question that is genuinely ambiguous, and a question this model happens
not to know. Telling these apart requires more than one opinion.

\paragraph{Measures from an ensemble.}
Suppose $M$ members produce distributions $p^{(1)},\dots,p^{(M)}$ for the same question. The
ensemble prediction is their average, $\bar{p} = \frac{1}{M}\sum_m p^{(m)}$, and two different
entropies can be formed from the same ingredients depending on whether entropy or averaging is
applied first. Taking the entropy \emph{of} the average gives the entropy of expected (EoE); taking
the average \emph{of} the entropies gives the expected entropy (EE). Entropy is concave, so the
first is never smaller than the second, and their difference is the mutual information between the
answer and the model
parameters~\citep{depeweg2018decomposition,malinin2018predictive,malinin2021uncertainty}:
\begin{equation}
  \underbrace{\mathcal{I}\big(y;\theta \mid x\big)}_{\text{knowledge uncertainty}}
  \;=\;
  \underbrace{\mathcal{H}\!\left[\bar{p}\right]}_{\text{total uncertainty (EoE)}}
  \;-\;
  \underbrace{\frac{1}{M}\sum_{m=1}^{M}\mathcal{H}\!\left[p^{(m)}\right]}_{\text{data uncertainty (EE)}} .
  \label{eq:decomposition}
\end{equation}
This decomposition is why an ensemble is more informative than a single model. EoE reports how
unsure the system is overall. EE is the part that survives inspecting each member, and so
reflects genuine ambiguity in the question or the audio. Mutual information is the part caused
by members disagreeing. A question on which every member is confident but no two agree has low EE
and high mutual information, a pattern no single distribution can express, and a strong signal
that the model does not really know the answer. With one member the decomposition collapses
($\mathrm{EoE}=\mathrm{EE}$) and mutual information is zero.

We use five measures: negative confidence and entropy from a single pass, and EoE, EE and mutual
information from an ensemble. All entropies are in nats. Training $M$ separate music
audio-language models is infeasible; Section~\ref{sec:abstention} obtains the members of
Eq.~\eqref{eq:decomposition} from one pretrained model by perturbing its input.

\section{Multiple-choice musical question answering}
\label{sec:mcqa}

In multiple-choice musical question answering, the model is given a short recording, a question and
$K$ candidate answers, exactly one of which is correct. We use the prompt format shipped with the
benchmark (Figure~\ref{fig:prompt}): options are labeled A--D and the prompt ends on a cue that
leaves the model with nothing sensible to emit but an option label.

\begin{figure}[t]
  \centering
  \fbox{\begin{minipage}{0.88\linewidth}
    \footnotesize\ttfamily\raggedright
    \textlangle audio\textrangle\\[1pt]
    Question: What rhythm pattern do the digital drums follow?\\
    Options: (A) Four on the floor. (B) Off-beat syncopation. (C) Scat singing.
    (D) E-guitar playing a simple melody.\\
    The correct answer is:
  \end{minipage}}
  \caption{The prompt passed to the model. The trailing cue means the next token is an option
  label.}
  \label{fig:prompt}
\end{figure}

We read the answer from the first generated token~\citep{robinson2023mcp}. Scoring the full option
text would need $K$ passes and is biased by length; parsing free-form generation is fragile. A
single forward pass instead yields the logit vector at the first generated position; we keep only
the four option-label tokens \texttt{A}, \texttt{B}, \texttt{C}, \texttt{D}, softmax them, and
take $\arg\max_k p_k$. This always returns a valid in-set answer, costs one pass rather than $K$,
and supplies the categorical distribution that Section~\ref{sec:uq} needs. Renormalizing discards
mass outside the option set, which is correct for a closed-set task; the resulting predictions
agree with those obtained by parsing generated text.

\section{Abstention in multiple-choice question answering}
\label{sec:abstention}

The readout of Section~\ref{sec:mcqa} turns each question into a distribution $p$ over the four
options, and the measures of Section~\ref{sec:uq} turn that distribution into an uncertainty score
$\mathcal{U}(x)$. Given a threshold $\tau$, the system answers with $\arg\max_k p_k$ when
$\mathcal{U}(x) < \tau$ and abstains otherwise. Raising $\tau$ increases coverage at the cost of
reliability. Quality is therefore not a property of any single $\tau$ but of how well $\mathcal{U}$
orders questions by their chance of being wrong~\citep{elyaniv2010foundations,geifman2017selective},
which is what we measure in Section~\ref{sec:metrics}.

\paragraph{Baseline: entropy of a single pass.}
The natural baseline is to run the model once, in the option ordering the benchmark provides, and
use the entropy of the resulting distribution~\citep{hendrycks2017baseline}. This is free, the
distribution is already computed in order to answer, but a single distribution cannot separate an
ambiguous question from one the model simply does not know, and cannot detect an unstable answer.

\paragraph{Pseudo-ensembles.}
An ensemble would resolve this, but training several music audio-language models is out of reach.
We instead construct what we call a \emph{pseudo-ensemble}. Let $t_1,\dots,t_M$ be transformations
of the input which are \emph{answer-preserving}: applying $t_m$ to $x$ changes the surface form of
the query without changing which candidate answer is correct. Running the model once on each
transformed input gives $M$ distributions over the options,
\begin{equation}
  p^{(m)} \;=\; P\big(y \mid t_m(x), \theta\big), \qquad m = 1,\dots,M ,
  \label{eq:pseudo-ensemble}
\end{equation}
which we treat as the members of Eq.~\eqref{eq:decomposition}: their average is the prediction, and
EoE, EE and mutual information follow. Diversity that a conventional ensemble gets from different
parameters is here obtained from different views of the same input, following test-time
augmentation~\citep{ayhan2018testtime} and virtual ensembles~\citep{malinin2021gradient}. A change
that carries no meaning should leave a competent model's belief untouched, so disagreement says
that the answer rests on how the question was written rather than on the music. The cost is $M$
forward passes and no training.

\paragraph{Option shuffling.}
Our main construction takes $t_m$ to be a permutation $\pi_m$ of the candidate answers, so that
member $m$ sees the same question with the options listed in a different order under the labels
A--D. This is answer-preserving, and it is a good source of diversity because language models
carry position
biases~\citep{zheng2024selectors,pezeshkpour2024sensitivity,alzahrani2024benchmarks}. The
distribution for member $m$ is over label \emph{positions}, not answers, so before averaging we map
it back through $\pi_m^{-1}$ so that index $k$ names the same candidate in every member. Averaging
the aligned distributions also cancels position bias, which is why the ensemble is more accurate
than the default ordering. With $K=4$ there are $24$ possible orderings; we use $M$ of them,
including the default, so the baseline above is the first member alone.

\paragraph{Audio corruption.}
A second construction leaves the text alone and perturbs the recording with additive Gaussian
white noise, lossy compression and a small time shift. None of these changes the correct answer
at the mild severities we use, but each changes the audio embedding. A member that relies on a
fragile acoustic cue will move; one that has really heard the relevant musical property will not.
The option order is held at the default so the effect is not confounded with shuffling. We also
report a silent-audio diagnostic, which is not answer-preserving but measures how much of the
benchmark can be answered from the text alone.

\paragraph{Label swapping.}
A third construction changes the option symbols from A/B/C/D to 1/2/3/4, W/X/Y/Z or
i/ii/iii/iv, leaving the audio and the answer order unchanged. A model that has understood the
question should give the same answer regardless of what the options are called. The constrained
readout never asks the model what it would freely generate: we restrict the first-token logits to
whichever four label tokens the prompt uses. In practice TinyMU still puts mass on A/B/C/D,
having been trained heavily on that convention, so this ensemble is poorly calibrated.

Of the three, we focus on option shuffling. It needs no choice of severity or scheme, and, as
Section~\ref{sec:results} shows, it gives both the largest accuracy gain and the strongest
abstention signal.

\section{Experiments}

\subsection{Data}

We evaluate on MuChoMusic~\citep{weck2024muchomusic}: $1{,}187$ four-option questions over $644$
tracks from MusicCaps~\citep{agostinelli2023musiclm} and the Song Describer
Dataset~\citep{manco2023songdescriber}. Every question has a single correct answer, so the
constrained readout and the option orderings are well defined. Questions are split into
\emph{knowledge} and \emph{reasoning} subsets and cover $23$ musical dimensions. Accuracy on the
benchmark is widely reported, so our numbers are comparable with published results for much larger
models. A known weakness is that some questions can be answered from the text
alone~\citep{weck2024muchomusic}; the silent-audio diagnostic measures this.

\subsection{Model}

We use the released TinyMU checkpoint~\citep{li2026tinymu} ($229$M parameters) with no further
training. A compact model is the right test case: it is wrong more often, so abstention matters,
and each extra forward pass is cheap. Audio is loaded as mono at $32$\,kHz and truncated to the
first $10$ seconds; silence replaces the waveform with zeros of the same length. Predictions are
read from the first-token logits of Section~\ref{sec:mcqa} and are therefore deterministic, with
no dependence on sampling temperature. Each pseudo-ensemble uses $M=4$ members:
four option orderings including the default; the clean recording plus Gaussian noise
($\sigma=0.01$), lossy compression and a small time shift; or the label schemes A/B/C/D, 1/2/3/4,
W/X/Y/Z and i/ii/iii/iv.

\subsection{Evaluation metrics}
\label{sec:metrics}

We report accuracy, the fraction of the $1{,}187$ questions for which the constrained argmax
matches the annotated answer, overall and on the knowledge and reasoning subsets. Random guessing
sits at $25\%$. For a pseudo-ensemble the prediction is the argmax of $\bar{p}$.

Accuracy says nothing about abstention, so we assess each uncertainty measure with an error
retention curve. Questions are sorted by uncertainty and the most uncertain ones are removed first;
at each retention fraction $r$ we plot the error rate over the $r$ most confident questions. This
is the error a user would see if the system answered only the fraction $r$ of questions it was
most sure about and deferred the rest. A useful measure falls steeply, since the questions it
discards first are the ones the model gets wrong. An oracle that ranks every error first, and a
random ranking that stays flat at the base error rate, bound what is achievable. We summarize
each curve by its area (AUC-ERC), lower being better.

\section{Results}
\label{sec:results}

Table~\ref{tab:accuracy} reports accuracy on MuChoMusic. The upper block repeats published
figures from~\citet{li2026tinymu}; the lower block is our evaluation of the same checkpoint under
the constrained first-token readout of Section~\ref{sec:mcqa}.

\begin{table}[ht]
  \centering
  \caption{Accuracy (\%) on MuChoMusic. Published numbers above, our runs of TinyMU below.
  Dashes mark values that are not reported in the original papers.}
  \label{tab:accuracy}
  \small
  \begin{tabular}{llccc}
    \toprule
    Model & Size & Knowledge & Reasoning & All \\
    \midrule
    MusiLingo~\citep{deng2024musilingo}              & 7.1B  & 33.6 & 28.2 & 31.5 \\
    MU-LLaMA~\citep{liu2024mullama}                  & 7.7B  & 32.3 & 33.5 & 32.7 \\
    Mellow~\citep{deshmukh2025mellow}                & 167M  & 30.8 & 32.0 & 30.3 \\
    Audio Flamingo 2~\citep{ghosh2025audioflamingo2} & 4.4B  & --   & --   & 56.5 \\
    Audio Flamingo 3~\citep{ghosh2025audioflamingo3} & 8.3B  & --   & --   & 47.4 \\
    Qwen2-Audio-Instruct~\citep{chu2024qwen2audio}   & 8.4B  & 69.4 & 65.5 & 67.8 \\
    MiDashengLM~\citep{midashenglm2025}              & 8.3B  & --   & --   & 71.4 \\
    TinyMU~\citep{li2026tinymu}                      & 229M  & 58.3 & 59.6 & 58.6 \\
    \midrule
    \multicolumn{5}{l}{\emph{TinyMU, this work}} \\
    Silent audio                                     & 229M  & 49.2 & 53.0 & 50.7 \\
    Single pass ($M=1$)                              & 229M  & 56.1 & 56.8 & 55.7 \\
    Option shuffling ($M=4$)                         & 229M  & 58.2 & 61.9 & 59.2 \\
    Audio corruption ($M=4$)                         & 229M  & 56.9 & 58.7 & 57.2 \\
    Label swapping ($M=4$)                           & 229M  & 42.9 & 42.3 & 42.7 \\
    Option shuffling ($M=21$)                        & 229M  & 59.7 & 62.5 & 60.3 \\
    \bottomrule
  \end{tabular}
\end{table}

Our single-pass accuracy is $55.69\%$, below the published $58.6\%$. The gap is expected:
\citet{li2026tinymu} score free-form generations, whereas we read the first-token distribution and
truncate audio to $10$ seconds. The constrained readout is the right object for uncertainty
estimation, so $55.69\%$ is the baseline below. Silent audio still scores $50.7\%$, well above
chance, confirming that a substantial fraction of MuChoMusic can be answered from the text
alone~\citep{weck2024muchomusic}. The five-point gap to the single-pass result is the contribution
of actually listening.

Option shuffling with $M=4$ raises accuracy to $59.22\%$, a gain of $3.5$ points and slightly
above the published TinyMU number. The gain is larger on reasoning questions ($56.8\%$ to
$61.9\%$) than on knowledge questions ($56.1\%$ to $58.2\%$). Audio corruption helps less
($57.20\%$). Averaging $M=21$ orderings reaches $60.3\%$.

\paragraph{Uncertainty ranking.}
Table~\ref{tab:retention} compares single-pass measures with those from the option-shuffling
ensemble. Negative confidence is $-\max_k p_k$ of the predicted class. Entropy of expected and
negative confidence use $\bar{p}$; expected entropy and mutual information need more than one
member.

\begin{table}[ht]
  \centering
  \caption{Area under the error retention curve (lower is better). Ensemble measures use
  option shuffling with $M=4$.}
  \label{tab:retention}
  \small
  \begin{tabular}{lc}
    \toprule
    Uncertainty measure & AUC-ERC ($\downarrow$) \\
    \midrule
    Baseline entropy (single pass) & 0.293 \\
    Negative confidence            & 0.264 \\
    Entropy of expected            & 0.264 \\
    Expected entropy               & 0.261 \\
    Mutual information             & 0.306 \\
    \midrule
    Oracle ranking                 & 0.098 \\
    Random ranking                 & 0.407 \\
    \bottomrule
  \end{tabular}
\end{table}

\begin{figure}[ht]
  \centering
  \begin{subfigure}[t]{0.32\textwidth}
    \centering
    \includegraphics[width=\linewidth]{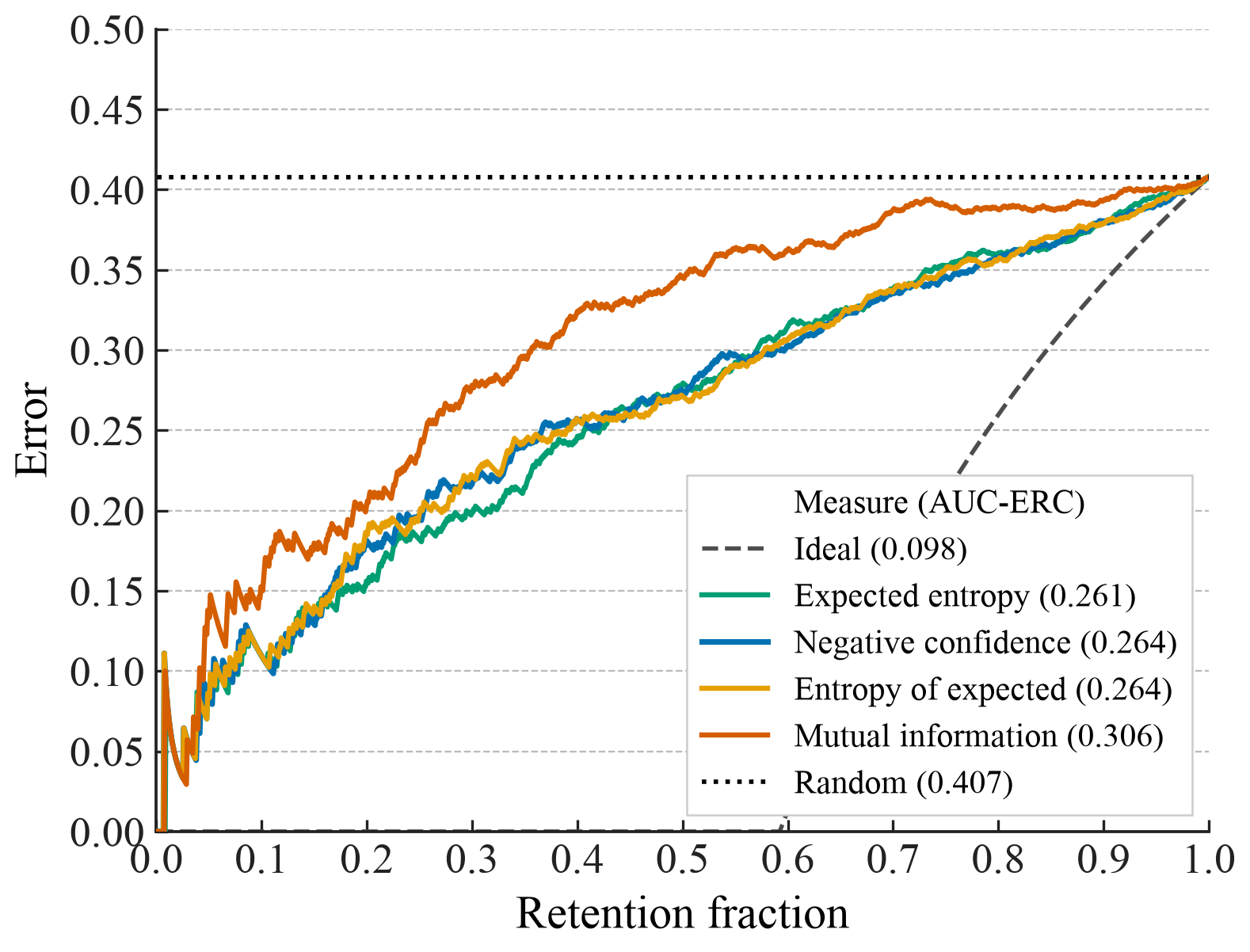}
    \caption{Option shuffling}
    \label{fig:erc-shuffling}
  \end{subfigure}
  \hfill
  \begin{subfigure}[t]{0.32\textwidth}
    \centering
    \includegraphics[width=\linewidth]{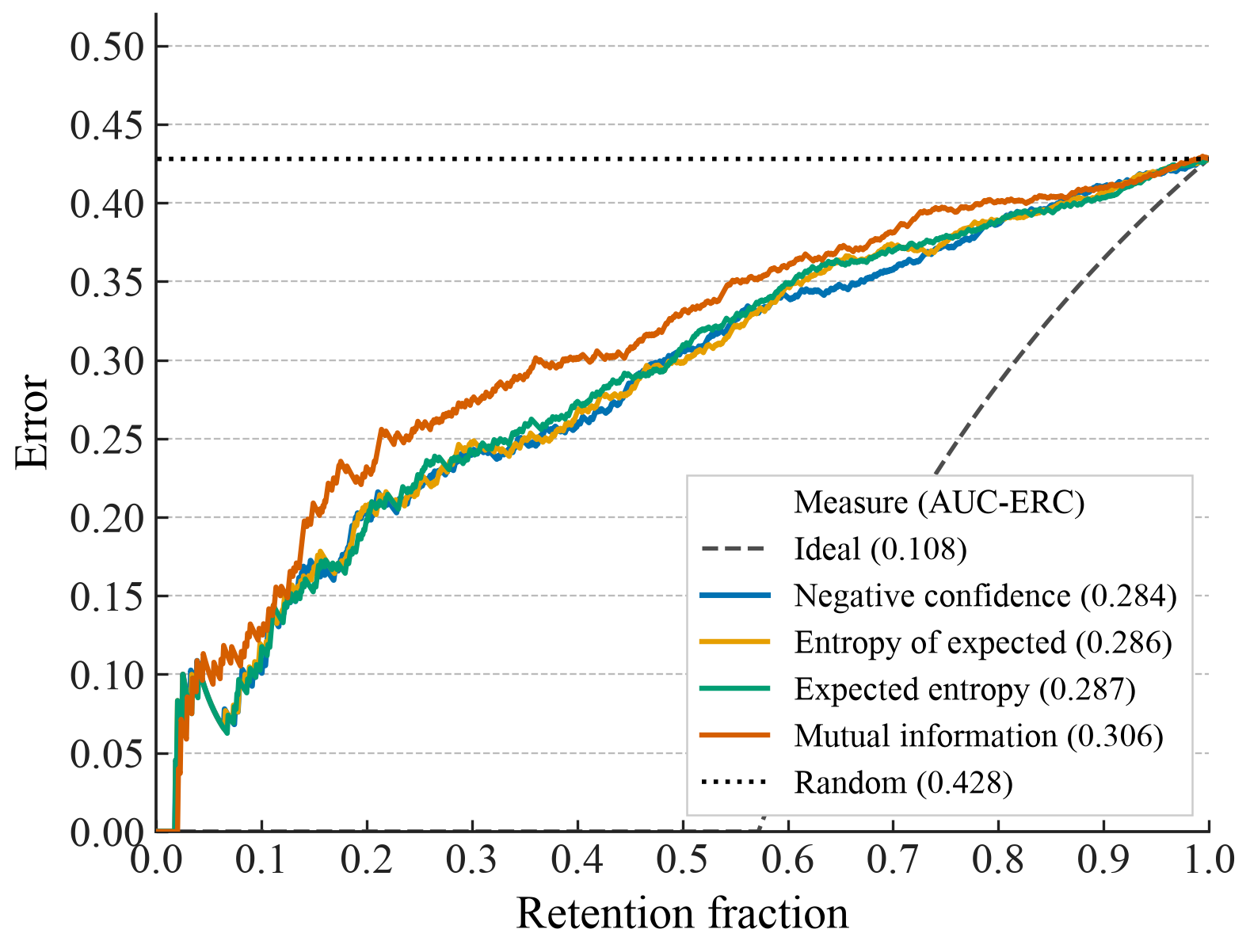}
    \caption{Audio corruption}
    \label{fig:erc-corruption}
  \end{subfigure}
  \hfill
  \begin{subfigure}[t]{0.32\textwidth}
    \centering
    \includegraphics[width=\linewidth]{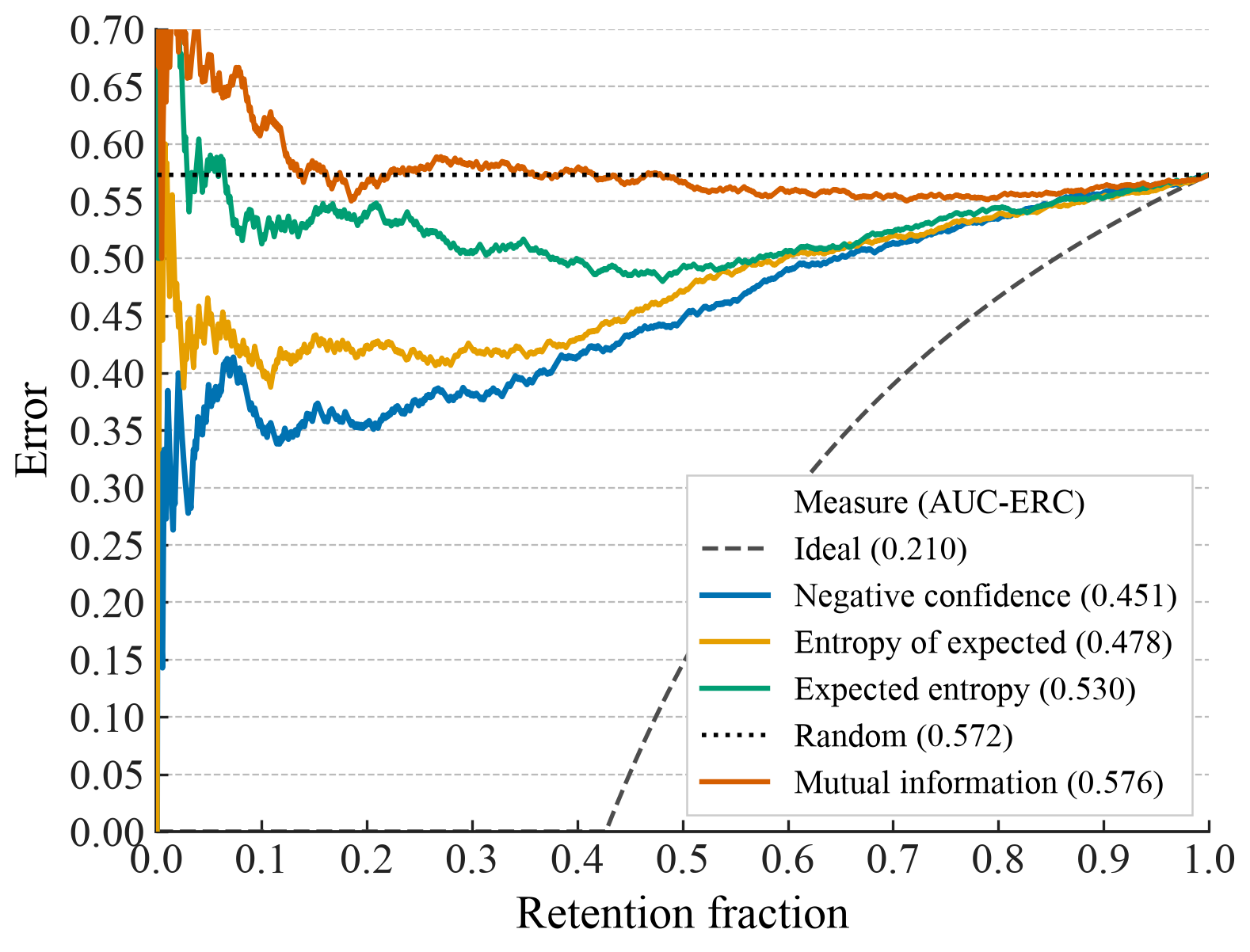}
    \caption{Label swapping}
    \label{fig:erc-labels}
  \end{subfigure}
  \caption{Error retention curves for each $M=4$ pseudo-ensemble, against the oracle and random
  references. A useful measure falls as the most uncertain questions are removed.}
  \label{fig:erc}
\end{figure}

Expected entropy is the strongest measure, reducing AUC-ERC from $0.293$ to $0.261$. Negative
confidence and entropy of expected both reach $0.264$. All three sit well below random ($0.407$)
and still above the oracle ($0.098$). Mutual information is worse than single-pass entropy
($0.306$): members often disagree on questions that $\bar{p}$ still answers correctly, because
option order moves mass between labels without changing the ensemble argmax. Mutual information
treats that disagreement as knowledge uncertainty and rejects questions the ensemble gets right.
Expected entropy keeps the residual uncertainty after averaging, which better matches whether the
final prediction is wrong. Figure~\ref{fig:erc} shows the corresponding curves.

\paragraph{Sources of diversity.}
Table~\ref{tab:sources} reports accuracy and every ensemble measure for the three constructions,
with $M=4$ throughout. The single-pass baseline is $55.69\%$ accuracy and $0.293$ AUC-ERC.

\begin{table}[ht]
  \centering
  \caption{Accuracy (\%) and AUC-ERC for each $M=4$ pseudo-ensemble. Lower AUC-ERC is
  better. The single-pass baseline is $55.69\%$ and $0.293$.}
  \label{tab:sources}
  \small
  \begin{tabular}{lccc}
    \toprule
    & Option shuffling & Audio corruption & Label swapping \\
    \midrule
    Accuracy ($\uparrow$)   & 59.22 & 57.20 & 42.70 \\
    \midrule
    Negative confidence     & 0.264 & 0.284 & 0.451 \\
    Entropy of expected     & 0.264 & 0.286 & 0.478 \\
    Expected entropy        & 0.261 & 0.287 & 0.530 \\
    Mutual information      & 0.306 & 0.306 & 0.576 \\
    \bottomrule
  \end{tabular}
\end{table}

Option shuffling is best on both accuracy and abstention. Changing the order of the answers is a
large, discrete perturbation, and language models are sensitive to
it~\citep{zheng2024selectors,pezeshkpour2024sensitivity}. Averaging the aligned distributions
cancels position bias, which is why accuracy rises, and leftover disagreement is informative about
questions the model has not settled. Audio corruption is weaker: a mild waveform change moves the
audio embedding less than a permutation moves the text, so the members stay closer together. It
still beats the single-pass baseline on both metrics. Label swapping does not help: TinyMU
continues to put mass on A/B/C/D even when the prompt uses another scheme, so the members are
poorly calibrated. Mutual information is the weakest measure in every column of
Table~\ref{tab:sources}, for the same reason as above: members disagree on questions that
$\bar{p}$ still answers correctly. In Figure~\ref{fig:erc}, expected entropy falls fastest under
option shuffling. Audio corruption is shallower but still well below a random ranking. Label
swapping stays high: even its best measure has a larger area than the single-pass baseline,
because the ensemble itself is much less accurate.

Relative to Table~\ref{tab:accuracy}, option shuffling at $M=4$ ($59.2\%$) exceeds Audio
Flamingo~2 ($56.5\%$) and the published TinyMU figure ($58.6\%$), while remaining more than
an order of magnitude smaller than the multi-billion-parameter systems. It does not close the
gap to Qwen2-Audio or MiDashengLM; that is the point of abstention. A compact model that
knows when to defer can still be useful next to a larger one. A practical reading of
Figure~\ref{fig:erc} is the error a user would see at a chosen coverage. No measure approaches
the oracle: even the best ranking leaves a residual of confidently wrong answers, which is why
we treat these scores as a filter rather than a calibrated probability of correctness.

\begin{figure}[!b]
  \centering
  \includegraphics[width=0.82\linewidth]{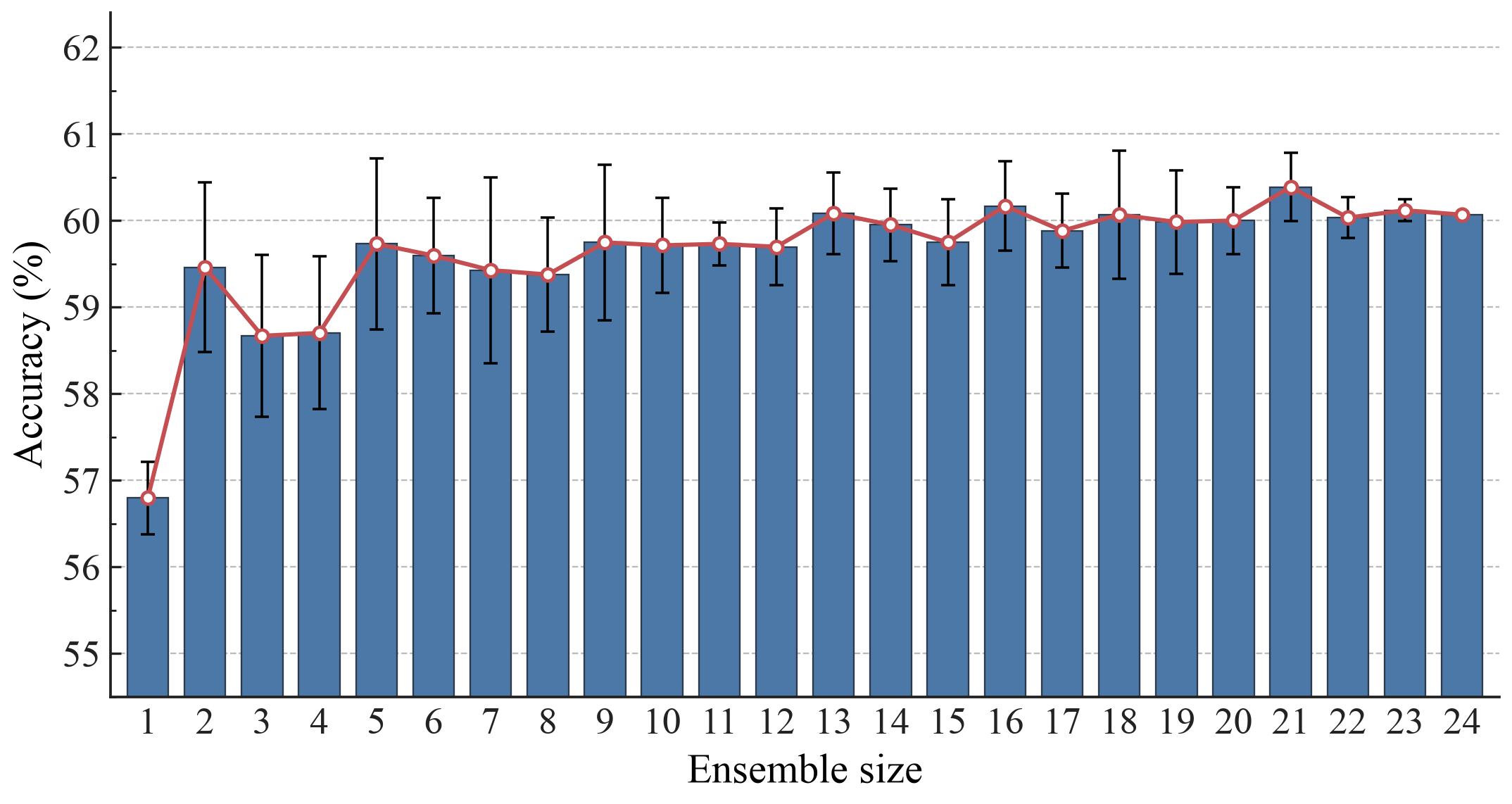}
  \caption{Option-shuffling accuracy versus ensemble size. For each $M<24$, five ensembles are
  drawn from the $24$ permutations; at $M=24$ there is a single complete ensemble.}
  \label{fig:ens-size}
\end{figure}

\paragraph{Ensemble size.}
Figure~\ref{fig:ens-size} varies the number of orderings. For each $M \in \{1,\dots,24\}$ we draw
five ensembles from the $24$ permutations (except $M=24$, where there is one complete ensemble)
and plot the mean accuracy with the range across draws. Accuracy rises with $M$. Most of the gain
appears in the first few members, already at $M=2$ the mean is close to $59\%$, and the mean
then continues to climb more slowly; the peak mean crosses $60.3\%$. The range across random
draws stays on the order of one point, so the $M=4$ setting used above is a cheap operating
point rather than a saturation point: a user who can afford more passes gets a further accuracy
gain, at a linear cost in compute.

\FloatBarrier
\section{Conclusion}
\label{sec:conclusion}

Music audio-language models are usually scored by accuracy on multiple-choice questions, which
treats a guess the same as a confident answer. We treat the same task as selective prediction.
Because training several copies of a music audio-language model is expensive, we build a
pseudo-ensemble from one released checkpoint by applying answer-preserving input changes.
Shuffling the candidate answers is the most effective of the three constructions we tried: it
raises accuracy from $55.7\%$ to $59.2\%$ with four orderings, and up to $60.3\%$ with
twenty-one, and it unlocks ensemble uncertainty measures that rank errors better than the
entropy of a single pass. Expected entropy is the best of those measures. The procedure uses a
few extra forward passes and no retraining, so abstention is cheap enough for a compact
on-device model. The extra passes also raise accuracy, so they are not spent on abstention alone.

\section{Limitations}
\label{sec:limitations}

The study uses one model and one benchmark. TinyMU is a good test case for a compact music
model, but we have not shown that the same ranking holds for multi-billion-parameter systems
or for other music QA sets. The constrained first-token readout assumes that the model binds
answers to option symbols; a model that ignores the labels would need a different scoring
rule. Option shuffling exploits a known position bias, so it may become less useful if future
models become invariant to option order, and it applies only to multiple-choice prompts.
Audio corruption depends on a choice of severity, which we did not tune extensively. Label
swapping failed on TinyMU because the model is tied to A/B/C/D labels; that finding may not
transfer. We report results from a deterministic readout of a single checkpoint, so the main
tables have no error bars; the only repeated sampling is the five random ensembles in the
size plot. Better uncertainty ranking is not the same as calibration: a low expected entropy means the
prediction is more likely to be correct, not that $\bar{p}$ is a well-calibrated probability.

A reliable abstention rule makes a music assistant more trustworthy when it answers, but it
can also hide systematic failures if the model is confidently wrong on a particular genre,
culture, or recording condition. We have not audited TinyMU for such gaps. The method itself
does not release a new model or dataset.

\FloatBarrier
\bibliographystyle{plainnat}
\bibliography{references}

\appendix
\FloatBarrier

\section{Implementation details}
\label{app:impl}

TinyMU~\citep{li2026tinymu} couples the self-supervised MATPAC++ audio
encoder~\citep{quelennec2025matpacpp} to SmolLM2-135M~\citep{allal2025smollm2} through a two-layer
projector. The encoder is frozen; the projector and language model are fine-tuned on
MusicSkills-3.5M. We use the released checkpoint without further training.

Audio is loaded as mono at $32$\,kHz and truncated to the first $10$ seconds. The silent-audio
diagnostic replaces the waveform with zeros of the same length and keeps the default option
order. For option shuffling, member distributions are aligned back to the original candidate
answers before averaging. For audio corruption the option order is held at the default; the four
members are the clean recording, additive Gaussian white noise with $\sigma = 0.01$, lossy
compression, and a small time shift. For label swapping the four schemes are A/B/C/D, 1/2/3/4,
W/X/Y/Z and i/ii/iii/iv, and the first-token softmax is taken over whichever four label tokens
the prompt uses.

All assets are public research releases: TinyMU~\citep{li2026tinymu},
MuChoMusic~\citep{weck2024muchomusic}, MusicCaps~\citep{agostinelli2023musiclm} and the Song
Describer Dataset~\citep{manco2023songdescriber}. Evaluation is inference only. The main $M=4$
setting is four forward passes per question; the $M=21$ and ensemble-size plots cost more passes
in proportion to $M$.

\end{document}